# In-situ nitriding of $Fe_2VAl$ during laser surface remelting to manipulate microstructure and crystalline defects

Leonie Gomell,*[a] Shao-Pu Tsai[a], Moritz Roscher[a], Ruben Bueno Villoro[a], Peter Konijnenberg[a,b], Stefan Zaefferer [a], Christina Scheu[a], and Baptiste Gault *[a,c]

Tailoring the physical properties of complex materials for targeted applications requires optimizing the microstructure and crystalline defects that influence electrical and thermal transport, and mechanical properties. Laser surface remelting can be used to modify the sub-surface microstructure of bulk materials and hence manipulate their properties locally. Here, we introduce an approach to perform remelting in a reactive nitrogen atmosphere, in order to form nitrides and induce segregation of nitrogen to structural defects. These defects arise from the fast solidification of the full-Heusler $Fe_2VAl$ compound that is a promising thermoelectric material. Advanced scanning electron microscopy, including electron channelling contrast imaging and three-dimensional electron backscatter diffraction, is complemented by atom probe tomography to study the distribution of crystalline defects and their local chemical composition. We reveal a high density of dislocations, which are stable due to their character as geometrically necessary dislocations. At these dislocations and low-angle grain boundaries, we observe segregation of nitrogen and vanadium, which can be enhanced by repeated remelting in nitrogen atmosphere. We propose that this approach can be generalized to other additive manufacturing processes to promote local segregation and precipitation states, thereby manipulating physical properties.

## Introduction

The microstructure of materials impacts their properties. For example, mechanical properties can be manipulated by segregation to grain boundaries [1–4]. For functional materials, like thermoelectric (TE) materials, microstructural engineering was shown to be effective in modifying transport properties and thereby increasing the conversion performance [5–7]. A defect-rich microstructure can lead to a reduction in thermal conductivity [8–11], which increases the TE performance. To fully understand the impact of the microstructure, it needs to be investigated from near-atomic to millimetres length scales.

Additive manufacturing has been rising in prominence over the past decades to enable the fabrication of complex parts with near net-shape and in three dimensions, for instance through selective laser melting (SLM) [12]. The particular thermal history of the build made by SLM has been reported to generate complex microstructures [13–15]. Further, it offers opportunities for designing alloys and processes to provide materials with unprecedented combinations of properties [16,17]. These benefits also have the potential to improve the performance of TE materials [18,19]. By adjusting the processing parameters, the microstructure, and hence properties, can be manipulated [20].

Of all the TE materials, we selected the promising full-Heusler $Fe_2VAl$ compound. To gain insights into the as-built microstructures of SLM processed $Fe_2VAl$, while avoiding the complexity the building process would create, we recently used laser surface remelting (LSR) to locally manipulate the microstructure of casted $Fe_2VAl$ [21]. Local measurements of the transport properties were correlated with advanced microstructural analysis to discuss the influence of structural defects [22]. Indeed, while some of these defects are easily visible using optical or scanning electron microscopy (SEM), others require near-atomic-scale analyses, which can be performed by atom probe tomography (APT). For $Fe_2VAl$, we showed vanadium nitride precipitates and/or segregation of V and N to dislocations, phase, and grain boundaries, which lead to an increase in TE performance [22,23]. Vanadium nitrides act as phonon scattering centres, reducing the thermal conductivity, which in turn increases the figure of merit [22]. The nitrogen in these earlier studies was an impurity uncontrollably introduced during the processing of the bulk samples.

The route for microstructural manipulations by using *in-situ* reactions during SLM or LSR processing has been overlooked. Yet during the remelting process, the material interacts with the surrounding atmosphere. Oxygen pick-up is a notorious issue that leads to unwanted oxidization or to O contents outside the specifications for safe usage of e.g. Ti-alloys [24,25]. Contrarily, *in-situ* nitriding may help to increase the performance of the material, both in terms of TE and mechanical properties [26–29].

Here, we show that the nitrogen concentration at defects can be controlled by LSR. By changing the protective atmosphere in the processing chamber of the SLM machine from Ar to $N_2$, we show that nitrogen can react with vanadium during remelting. Upon multiple passes, we find an increased nitrogen concentration in the sample. If the performance of a TE generator depends on the microstructure, the stability of the microstructure is also essential, even if it is often overlooked. We performed heat-treatments *ex-situ* at 500 K for up to 100 h, and found that the dislocations and grain boundaries are stable due to their characteristic as geometrically necessary dislocations (GND). Microstructural manipulation by LSR, or even extended to SLM, is not limited to $Fe_2VAl$ but can be used to change the properties of other intermetallic compounds containing reactive metals in a reactive atmosphere in a similar manner.

## Experimental

We synthesized $Fe_2VAl$ by melting stoichiometric amounts of pure Fe (99.9%, Carboleg GmbH), Al (99.7%, Aluminium Norf GmbH), and V (99.9%, HMW Hauner GmbH) in an arc furnace. The sample was flipped over and remelted four times to ensure homogeneity. The average composition was obtained by inductively coupled plasma optical emission spectrometry (ICP-OES) to be $Fe_{50.07}V_{24.93}Al_{25.95}$ (at.%). The casted sample was ground down to 600 grit SiC paper to ensure a homogeneous surface for LSR.

The ytterbium-fiber laser used has a wavelength of 1070 nm and is focused to a spot size of 90 µm. The laser scanning speed was 1400 mm/s and the laser power was set to 200 W. The scanning strategy consists of single lines, which are separated by at least 0.5 mm. Between the remelting of the different lines, the sample was allowed to cool down to room temperature by pausing for 90 seconds. Some lines were remelted 3× and 5× on the same track. The remelting process was performed in an inert argon atmosphere or a reactive $N_2$ atmosphere to induce *in-situ* nitriding. In the following, these samples will be called $LSR_{Ar}$ and $LSR_N$, respectively. The residual oxygen concentration was below 80 ppm.

Microstructural investigations using SEM, including electron backscattered diffraction (EBSD), energy-dispersive X-ray spectroscopy (EDX), and electron channelling contrast imaging (ECCI) were performed in cross-sectional view and top view. Prior to the microscopy experiments, the samples were polished down to 0.05 µm colloidal silica. In top view, these samples were imaged without polishing and after grinding of circa 5 µm and 20 µm, similar to the samples remelted in Ar atmosphere presented in Ref. [22]. These samples will be called $top_0$, $top_5$, and $top_{20}$ in the following. The $top_5$ sample is also used for investigations of the stability of defects. The samples were held at 500 K for 2 hours before a second investigation and additionally for 5 days before a third investigation of the microstructure. The temperature was chosen in agreement with the peak $zT$ temperature of thermoelectric $Fe_2VAl$ [30,31]. To avoid oxidation during the annealing process, the annealing chamber was flooded with Ar gas and a Ta getter was used.

For SEM imaging in backscattered electron (BSE) mode and ECCI, a Zeiss SEM 450 was used. EBSD and EDX were performed in a Zeiss Sigma, equipped with an EDAX OIM EBSD System with a Hikari camera. The acceleration voltage was set to 15 kV for EDX, 20 kV for EBSD, and 30 kV for ECCI. ECCI was used for direct observation of dislocations and other defects and for calculating the dislocation density. 2D EBSD mapping was conducted with a step size of 200 nm on a hexagonal grid. Furthermore, large-volume 3D EBSD was performed using the in-house built automated system ELAVO 3D [32], which is based on automated serial polishing combined with automated EBSD mapping. Controlled polishing is performed with a QATM Saphir X-Change auto-polishing machine. On the present material, a removal rate of approx. 1 µm was achieved by polishing with 1 µm diamond suspension for 80 seconds, 50 nm oxide suspension polishing for 360 seconds, and sufficient cleaning. The removal rate was determined by an M-shaped pattern applied to a sample side perpendicular to the surface of interest using an FEI Helios plasma focused-ion beam (PFIB). The used parameters are summarized in Tab. S1 (ESI) and the removal rate is displayed in Fig. S1 (ESI). A Universal Robots UR5 transfers the sample between a Zeiss Crossbeam 1540 XB SEM and the polishing station. For 3D EBSD, a cubic measurement raster with a voxel size of 1 µm³ was chosen. Each volume contains 100 slices, translating into a measured depth of 100 µm. Due to the different sizes of the melt pools, the analysed volume varies between approx. 700,000 µm³ and 1,900,000 µm³ for the 1× and 5× remelted samples, respectively. An in-house version (2.0.22.2000) of the QUBE software package was used for post-processing of all 3D-datasets [33].

Near atomic-scale investigations were performed by APT. A dual-beam focused ion beam instrument (FEI Helios Nanolab 600/600i), equipped with a $Ga^+$ ion source, was used to prepare needle-shaped specimens. The lift-out was prepared from the cross section at a position approximately 5 µm below the surface. The procedure is described in Ref. [34]. APT was conducted using a LEAP 5000 XS instrument (Cameca Instruments) operated in laser pulsing mode. The pulse energy was set to 50 pJ, the pulse repetition rate to 200 kHz, and the detection rate to 2 %. The base temperature was kept at 60 K. The data was reconstructed and analysed by AP Suite 6.1 (Cameca Instruments).

## Results and Discussion

Fig. 1 shows BSE images of the cross-section of 1×, 3×, and 5× remelted melt pools of the $LSR_{Ar}$ and $LSR_N$ samples (Fig. 1a and b, respectively). The position with regard to the melt pool is sketched next to the images. BSE images in top view are displayed in Fig. S2 (ESI). The atmosphere does not influence the microstructure on a scale noticeable by BSE, EBSD, and EDX (Figs. 1, 3, S3, S7, S8 in ESI). Hence, the higher thermal conductivity of $N_2$ compared to Ar [35], which was shown to change melt pool dimensions [29], can be neglected here. The thermal conductivity of the material is three orders of magnitude higher than that of the surrounding gases. Hence, the heat transport and solidification processes are guided by the material's properties and the influence of the gases are neglectable. This enables the use of the same processing parameters (i.e. laser scanning speed, laser power, etc.) without the need to optimize the parameters again after changing the atmosphere. Optimizing the processing parameters (for example to reduce cracking or to adjust the melt pool size and grain size) is often a time-consuming task. Hence, changing the properties of a material by changing the atmosphere can be more easily applied than changing the other process parameters.

All micrographs show elongated solidification cells, i.e. individual grains, which grow in a columnar fashion (Figs. 1, S2) from the substrate along the maximum thermal gradient directions. For all remelting conditions, cracks are observed, which appear at, or close to, high-angle grain boundaries (HAGBs) of the underlying casted sample. The melt pool and grain sizes are given in Table S2 and discussed in the supplementary information. In short, while the melt pool size increases for repeated remelting, the grain size decreases. Further, the melt pool is destabilized and exhibits fluctuations in size and shape after repeated remelting. While all 1× remelted regions look similar, there are strong fluctuations for 3× and 5× remelted regions. This is the main reason for the increased standard deviations of the melt pool sizes, which are summarized in Table S2. The 3D EBSD reconstructions (Fig. 2 and videos), which include 100 µm of each molten track, similarly show a strongly fluctuating melt pool size for the 5× remelted melt pool. This indicates that multiple remelting facilitates melt pool instabilities, which offers possibilities for even more complex microstructures than 1× remelting.

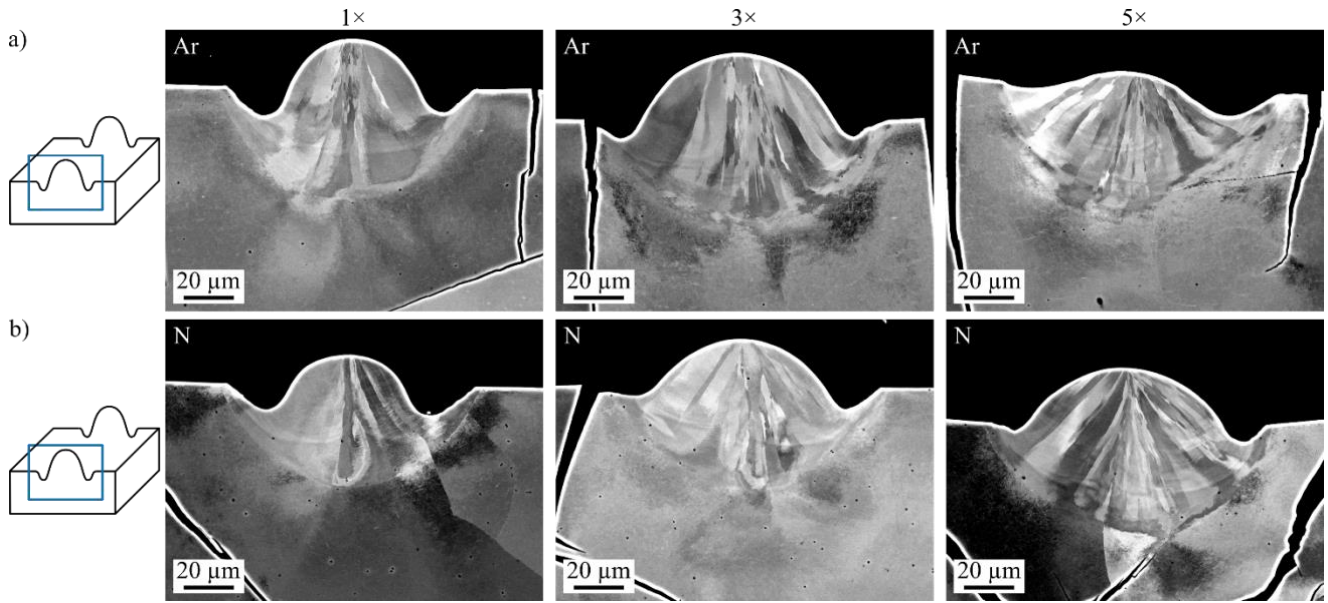

**Figure 1**: BSE images of the cross-section of a) the LSR$_{Ar}$ sample, b) the LSR$_N$ sample. The left, middle, and right columns show 1×, 3×, and 5× remelted regions.

Figs. 2 and 3 show the 3D and 2D EBSD evaluations of the LSR$_{Ar}$ sample, respectively. The 2D EBSD map is taken after finishing the 3D procedure at the last slice with a higher resolution. This enables us to analyse the grain structure and misorientations across grain boundaries in more detail. EBSD maps from the LSR$_N$ samples in cross-sectional and top view are given in the supplementary material (Fig. S3 and S4, ESI). Fig. 2a, b and Fig. 3a show the inverse pole figure (IPF) map along laser scanning direction. For the 3D scan, different viewing directions are shown. In the melt pool region, newly solidified grains adopt almost the same orientation as the substrate, which indicates an epitaxial growth mechanism. This can be confirmed by almost the same colour coding (orange). However, slight colour contrast is observed in both, the heat-affected zones and melt pool interiors, which is possibly due to strains caused by solidification. Within the 1× remelted melt pool, a grain boundary of the casted grain is observed, indicated by the colour contrast in the IPF. Its shape within the melt pool is discussed in more detail in the supplementary materials (Fig. S5).

No newly formed HAGBs are observed in the melt pool, also not for repeatedly remelted regions, except for grains close to a crack in the 3× remelted region. 18 grains with a misorientation higher than 5° with regards to the substrate grain are found. These grains are shown in supplementary Fig. S6. One of these grains is also easily visible in Figs. 2a, b and 3a (highlighted by the dashed circle), and movie sequence 2, as it appears blue in the IPF map, indicating a normal close to (111) in sample $z$-direction. We attribute this growth behaviour to the missing substrate grain for epitaxial growth. The crack was formed before the last remelting and the melt was not able to close the crack before solidifying. Hence, instead of epitaxial growth, new grains nucleated with a random orientation. These grains then grew in the direction of the thermal gradient until they reached the surface.

The grain boundaries within the melt pool can be characterized by their misorientation. For 3D EBSD, Fig. 2c shows the local average disorientation (LAD). In 2D, a kernel average misorientation (KAM) map (Fig. 3b) and a grain reference orientation deviation (GROD) map (Fig. 3c) are used for visualization. The LAD and KAM maps show the average misorientation with regard to the next nearest neighbour in the kernel, while the GROD map indicates the misorientation of each point with regard to the mean grain orientation of the substrate as reference. Most grain boundaries show a misorientation angle of 1° - 2°. Compared to the substrate grain, the maximum misorientation is approx. 10°. No clear trend of the misorientation could be observed for the multiple remelted regions. Due to the misorientations within the grains, dislocations are created to compensate for the strains generated. The KAM data is also used to calculate the density of the GNDs, as described in the supplementary information. On average, a GND density of $1.2 \cdot 10^{13}$ m$^{-2}$ is observed. In order to explain the formation of GNDs inside grains in melt pools, GROD axis analyses were carried out. However, the crystal rotations do not follow any clear or systematic trend. Further analyses must be conducted to understand the origin of crystal rotations inside each grain fully. Besides this, the 3D EBSD data show that there is no significant evolution of microstructure along the scan track, i.e. most of the data can be well-interpreted by looking on 2-dimensional cross sections.

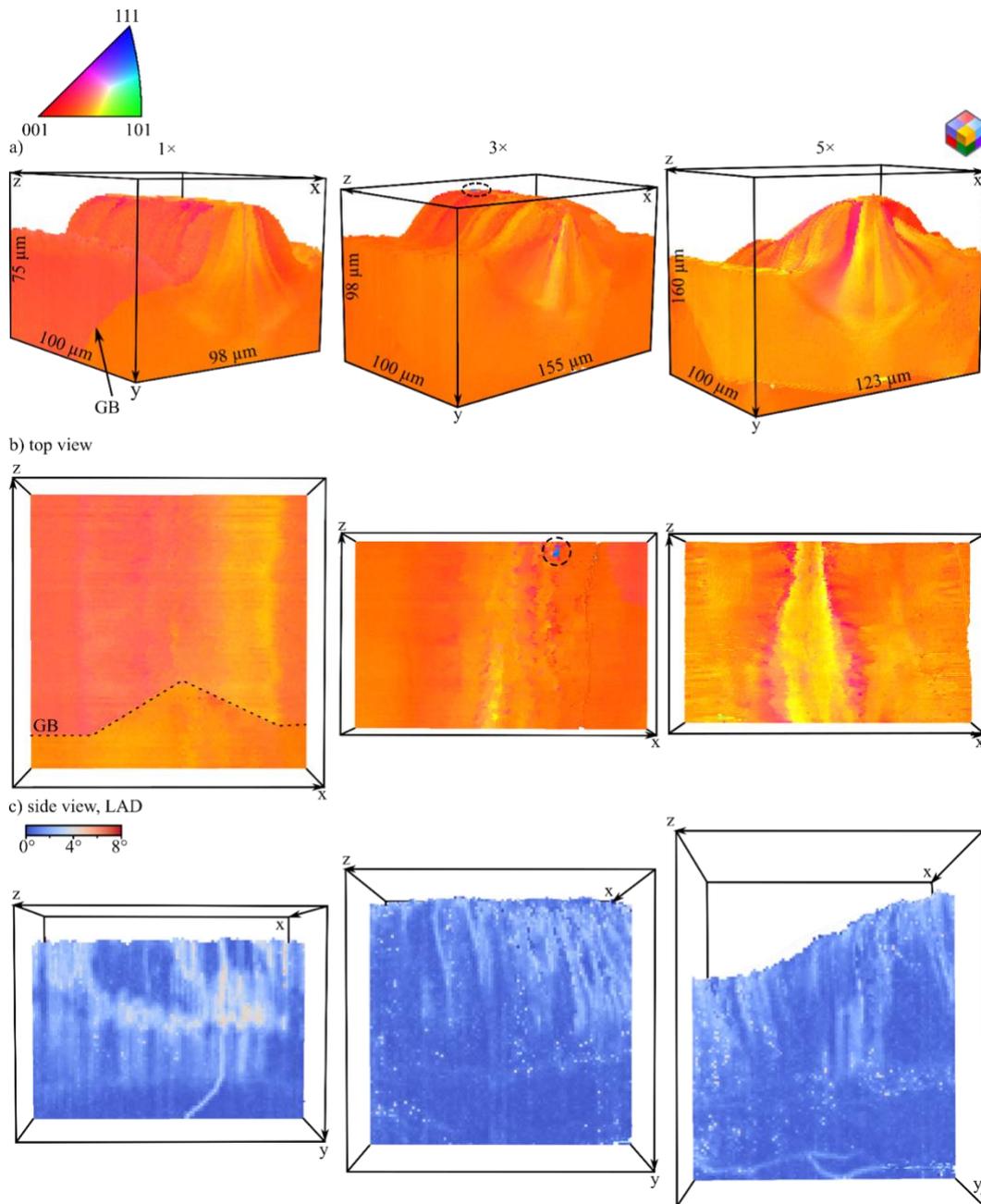

**Figure 2**: Large-volume 3D EBSD of the 1×, 3×, and 5× remelted tracks of the LSR$_{Ar}$ sample. a, b) IPF, shown in different viewing directions. The dashed line marks the GB of the substrate, and the dashed circle marks the non-epitaxially grown grain. c) LAD plot in side view, showing the LAGBs in the melt pools.

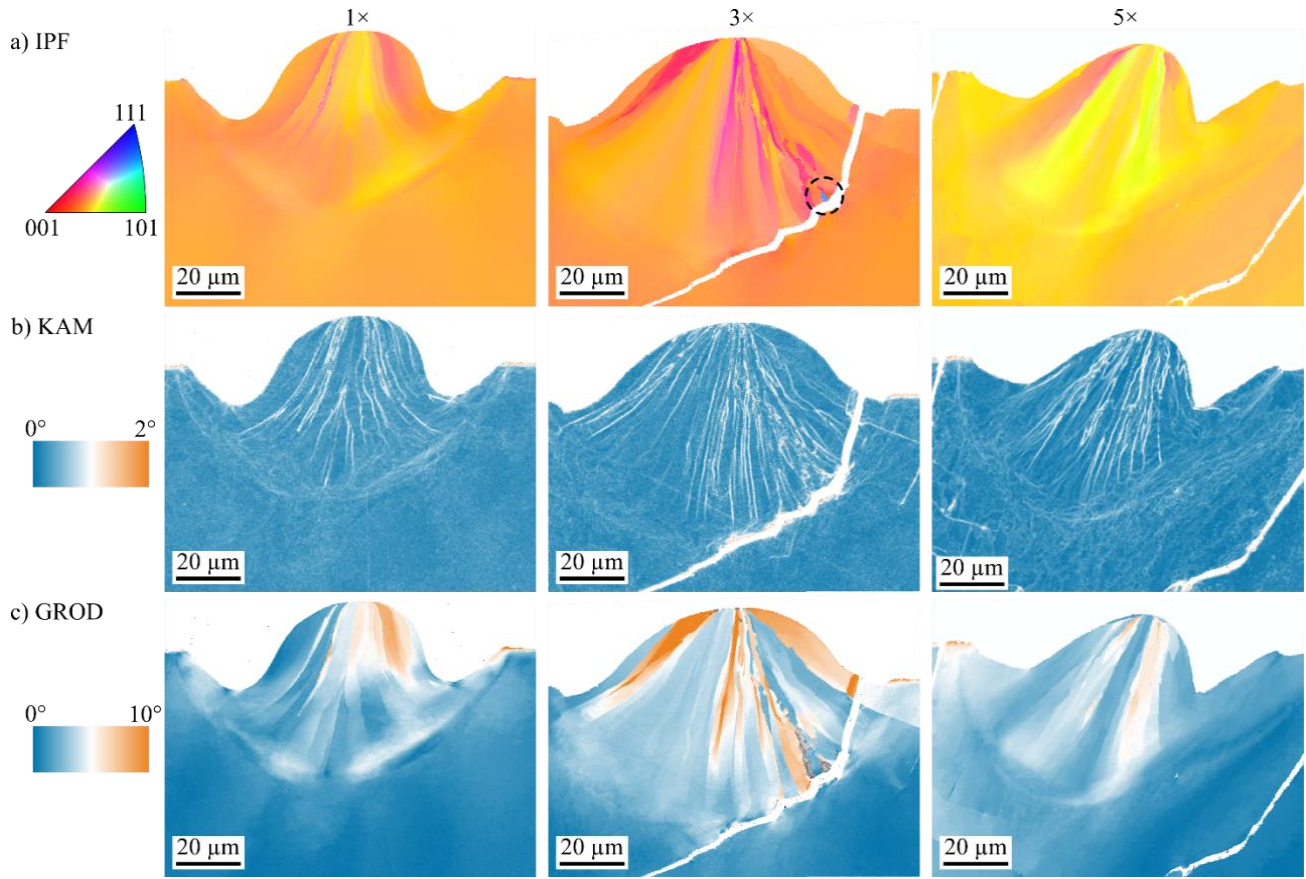

**Figure 3:** 2D EBSD maps of the LSR$_{Ar}$ sample after 1×, 3×, and 5× remelting. a) IPF maps along sample normal direction, which is parallel to the laser scanning direction. b) KAM maps, showing the misorientation between second nearest neighbors on a scale from 0° to 2°. c) GROD maps, showing the misorientation with regard to the substrate grain on a scale from 0° to 10°.

## Scanning Electron Microscopy – Electron Channelling Contrast Imaging (ECCI)

When the unpolished sample surface is analysed more closely using ECCI, nano-platelets are observed in the surface layer, which exclusively grow in N$_2$ but not in Ar atmosphere. Fig. 4a shows the surface of the LSR$_{Ar,1×}$ sample. Dislocations with a density of about $1 \cdot 10^{14}$ m$^{-2}$ are observed. The dislocations are aligned in a preferred direction, approximately parallel to the laser scanning direction. On the other hand, the surface of the LSR$_N$ sample (Fig. 4b) does not show any dislocations, but platelet-shaped precipitates, with an average length of $49 \pm 14$ nm, an average width of a few nm, and a number density of approx. $3 \cdot 10^{14}$ m$^{-2}$. Similar precipitates can be found in the 3× and 5× remelted melt pools (Fig. S9, ESI). Since these precipitates only grow in N$_2$ atmosphere and the previously observed tendency of V to form VN$_x$ [23,36,37], we may assume that these precipitates are VN$_x$. Controlled ECCI (i.e. ECCI under well-controlled Bragg-diffraction) was conducted on the 5× remelted melt pool (inset in Fig. S9, ESI), indicating that the precipitates lie on the (001) lattice plane of the host matrix, which is consistent with the 90° angle found between individual particles visible in Fig. 4b.

The precipitates only appear in a surface layer of the melt pool, which stays longer in contact with the N$_2$ atmosphere than the interior of the melt pool and is the last part to solidify. During solidification, most of the impurities remain in the melt and are expelled from the solid, similar to the typical Scheil model for solidification. The low solubility of N in Fe$_2$VAl was observed by APT previously [22,23,37] and will be shown again below: nitrogen segregates to dislocations and grain boundaries and is hardly found in the matrix (<0.1 at.%). The remaining melt enriches in N and gets increasingly supersaturated, facilitating the formation of precipitates in the steps of solidification. Additional reactions of the N$_2$ atmosphere and the solid at elevated temperatures might lead to further integration of N in the material.

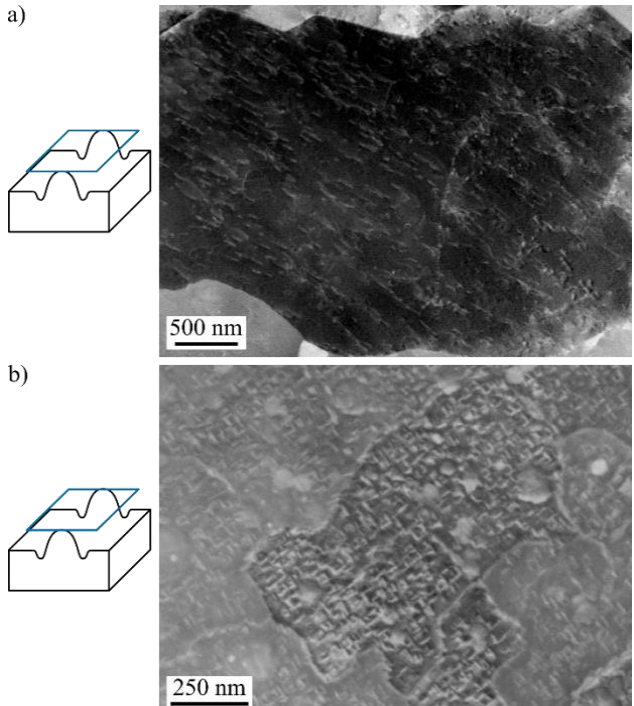

**Figure 4**: ECC images of the unpolished $top_0$ sample, a) $LSR_{Ar,1\times}$, and b) $LSR_{N,1\times}$. The $LSR_{Ar}$ sample shows dislocations, while the $LSR_N$ sample shows platelet-shaped nano-precipitates.

If the top part is polished away and the $top_5$ sample is observed, only dislocations but no precipitates can be found in both, the $LSR_{Ar}$ and $LSR_N$ samples (Fig. 5). The dislocation density is on the order of $3 \cdot 10^{13}$ m$^{-2}$, which is slightly lower than that on the unpolished surface remelted in Ar atmosphere. No significant difference in dislocation density occurs between $LSR_{Ar}$ or $LSR_N$ samples or for the multiple times remelted regions. This density agrees well with the calculated GND density of $1 \cdot 10^{13}$ m$^{-2}$, and hence, the dislocations found in ECCI can be classified as GNDs (i.e. they all have similar Burgers vectors).

GNDs are essential to accommodate the lattice rotations and can only annihilate by recrystallization at high temperatures. This is an advantage over other dislocations, which could migrate and annihilate at the working temperatures of a TE device, i.e. at 500 K for $Fe_2VAl$. To show that the dislocations found in the melt pool are stable, the same position is imaged before and after heat treatment for 2 hours and additionally, 5 days. Fig. 5 shows the $LSR_{Ar}$ and $LSR_N$ samples remelted once. The 3× and 5× remelted samples are shown in the supplemental materials (Fig. S10, ESI). By comparing the position and density of the dislocations and grain boundaries, we show evidence that they are stable. However, this experimental design has some drawbacks. Mirror forces might pin the observed dislocations to the surface, and dislocations within the bulk could behave differently and migrate. Yet, due to their character as GNDs, we can confidentially claim that the microstructure generated by LSR is stable at the temperature of highest TE performance of $Fe_2VAl$. To finalize the ECCI investigations, Fig. S11 (ESI) shows micrographs of grain boundaries. Single dislocations can be seen, confirming the low-angle grain boundary (LAGB) character observed by EBSD. The distance between the dislocations can be used to calculate the misorientation angle (ESI).

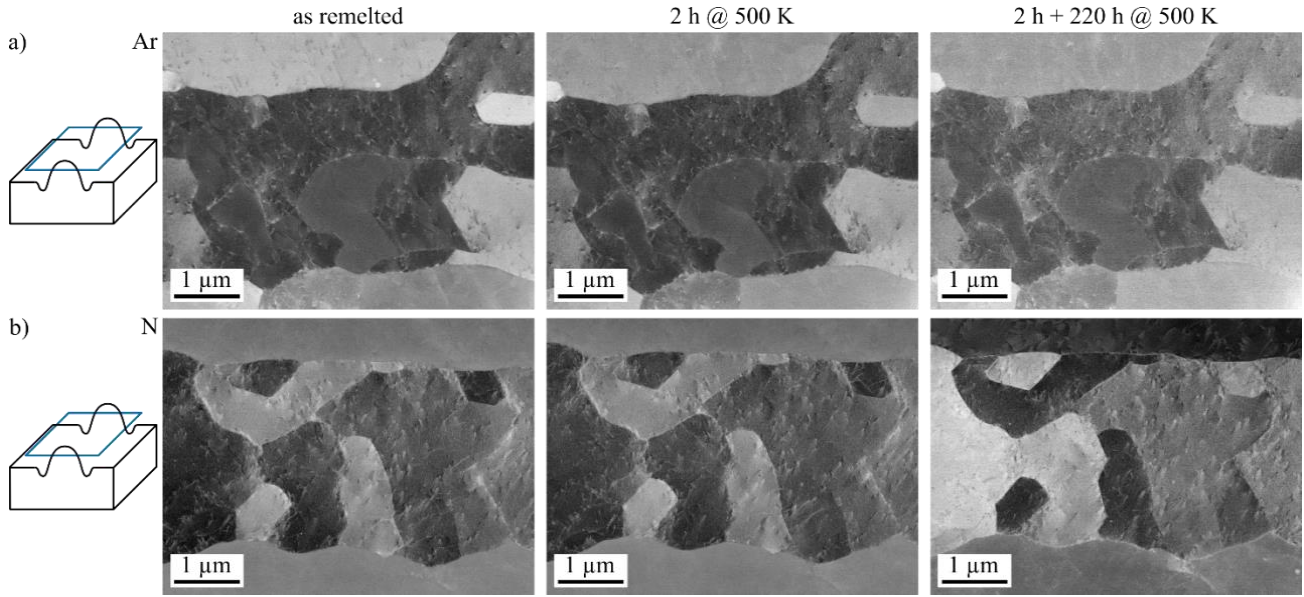

**Figure 5**: ECC images of the top$_5$ sample, a) LSR$_{Ar,1\times}$ and b) LSR$_{N,1\times}$. The positions of the grain boundaries and dislocations do not change after heat treatment, indicating a stable microstructure. Contrast variations in the images are due to different channeling conditions.

Atom Probe Tomography – Nanoscale Compositional Analysis

Using SEM methods, no indication of nitriding was observed within the melt pool interior, only on the surface. APT is then used to quantify the composition on the nanometre scale. The APT specimens were taken approx. 5 µm below the surface, and hence, the precipitates shown in Fig. 4 were not observed. 1×, 3×, and 5× remelted melt pools of the LSR$_{Ar}$ and LSR$_N$ samples are compared.

In a first step, the composition of the APT specimens is analysed and given in Tab. 1. A constant and small amount of N is found in the LSR$_{Ar}$ sample, which can be attributed to impurities of the raw material and to residual N$_2$ in the atmosphere during casting. The LSR$_{N,1\times}$ sample shows no increase in overall N concentration, but the N content is increased after repeated remelting. All nitrogen is detected as VN$^+$ or VN$^{2+}$ molecular ions typically detected in (carbo)nitride precipitates in Fe-based materials [38,39] and decomposed to obtain the N composition. Remelting further changes the composition: Al is preferentially lost during this process, as observed by APT analyses and EDX maps (Figs. S7 and S8). This loss is attributed to preferential spatter and evaporation of Al [22]. While Al is depleted, the V composition is increased slightly. Such an off-stoichiometric composition of Fe$_2$V$_{1-x}$Al$_{1+x}$ ($x<0$) was shown to increase the thermoelectric figure of merit up to 0.13 due to an optimized charge carrier concentration [30,40].

*Table 1*: Averaged composition of the different specimens determined by APT. For every sample, more than 200 million ions have been measured. The composition of the substrate, determined by ICP-OES, is given for comparison.

| Sample | Fe (at.%) | V (at.%) | Al (at.%) | N (at.%) |
|---|---|---|---|---|
| **1×, Ar** | 49.9 ± 0.1 | 26.3 ± 0.1 | 23.8 ± 0.1 | 0.02 ± 0.02 |
| **3×, Ar** | 50.2 ± 0.1 | 27.2 ± 0.1 | 22.6 ± 0.1 | 0.02 ± 0.02 |
| **5×, Ar** | 50.4 ± 0.1 | 28.1 ± 0.1 | 21.4 ± 0.1 | 0.03 ± 0.02 |
| **1×, N** | 50.0 ± 0.1 | 26.2 ± 0.1 | 23.9 ± 0.1 | 0.02 ± 0.02 |
| **3×, N** | 50.4 ± 0.1 | 26.8 ± 0.1 | 22.7 ± 0.1 | 0.04 ± 0.03 |
| **5×, N** | 50.2 ± 0.1 | 27.8 ± 0.1 | 22.0 ± 0.1 | 0.08 ± 0.03 |
| **Substrate (ICP-OES)** | 50.07 | 24.93 | 25.95 | 0.01 |

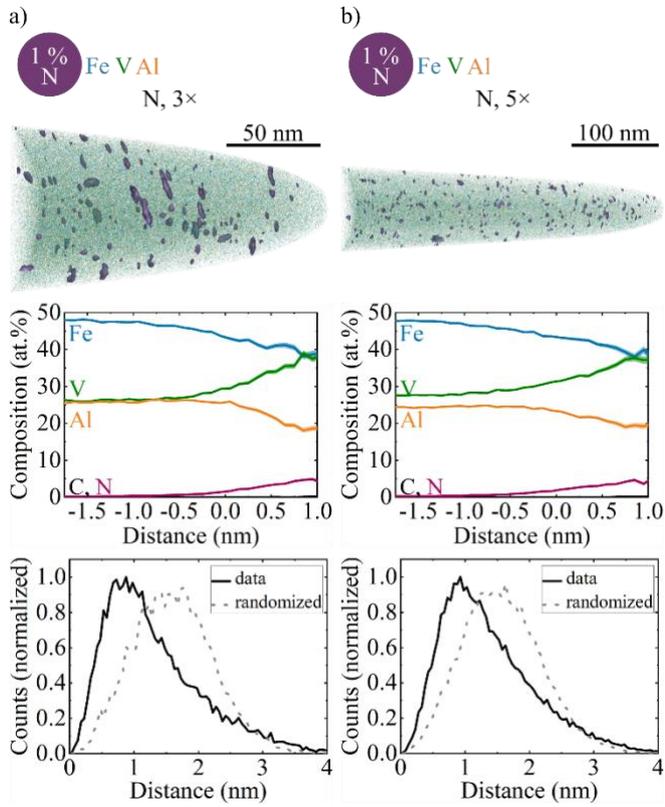

**Figure 6:** APT reconstruction of LSR$_N$ specimens after a) 3×, and b) 5× remelting. Clusters are found in both specimens, while they could not be observed in the other specimens. The proxigram shows the composition of the clusters, which are enriched in V and N and depleted in Fe and Al. The cluster count distribution plotted for VN molecular ions suggests significant clustering.

In a second step, the composition of clusters, dislocations, and LAGBs is analysed. These defects are known to introduce phonon scattering centres, which decrease the thermal conductivity, and hence, increase the thermoelectric performance [41–44]. VN$_x$ clusters, indicated by an iso-composition surface of 1 at.% N (Fig. 6), are only found in the sample remelted 3× and 5× in nitrogen atmosphere. As random fluctuations in the data set can also create very small clusters, only clusters larger than 3 nm³ are considered in the calculations. Fig. 6 shows the cluster composition, calculated as a proxigram, and the nearest neighbour distribution, indicating the non-random clustering of VN molecular ions. The proxigram shows the enrichment of V and N together with a depletion of Fe and Al. For both, the 3× and 5× remelted specimens, the maximum N concentration is approx. 8 at.%, an increase of two orders of magnitude compared to the average N concentration. The composition of the clusters was determined using 1D-compositional profiles of five individual clusters (not shown) to be Fe$_{27\pm5}$V$_{53\pm7}$Al$_{12\pm3}$N$_{8\pm1}$ (at.%). The impact of nanoparticles on the TE properties, especially the thermal conductivity, was discussed by Mingo et al. [41] on the example of silicide nanoparticles in SiGe. They showed a reduction in thermal conductivity due to phonon scattering at the nanoparticles. Hence, we anticipate that the clusters observed in the LSR$_N$ specimens will also contribute to a reduction of the thermal conductivity.

Dislocations, including GNDs, within the melt pool can scatter phonons by their strain field and core. In Ref. [22], we reported a decrease of the thermal conductivity after 1× remelting in Ar atmosphere, attributed to dislocation scattering. As core scattering depends on the presence of a Cottrell atmosphere [43,44], we assume that a change in the composition due to nitriding will further affect the thermal transport. The Cottrell atmosphere is observed by APT.

Dislocations have been found by APT in all samples, as shown in Fig. 7. However, in the LSR$_{N,3\times}$ and LSR$_{N,5\times}$ samples, dislocations are easier to observe and visualize due to more pronounced segregation. Iso-composition surfaces of N are superimposed on the reconstructions, with the used concentration threshold given next to the reconstructions in Fig. 7. Randomly distributed clusters of iso-composition surfaces, not belonging to dislocations, are shown in transparent grey for better visualization of the features of interest. The LSR$_{Ar}$ samples show little segregation of N towards the dislocations with a maximum concentration of approx. 0.5 at.%. It appears, similar to the cluster formation but with a less strong level, that nitrogen is co-segregated by vanadium. Again, a depletion of Fe and Al is observed. The segregation of V and N can be increased by changing the atmosphere to N$_2$. The maximum N concentration at dislocations is approx. 0.5 at.%, 2 at.%, and 2.5 at.%, for the 1×, 3×, and 5× remelted regions, respectively. Therefore, segregation could be increased by an order of magnitude compared to the

LSR$_{Ar}$ sample. In addition, also the V segregation and the depletion of Fe and Al is more pronounced. We attribute this to co-segregation of V and N, triggered by the higher N concentration.

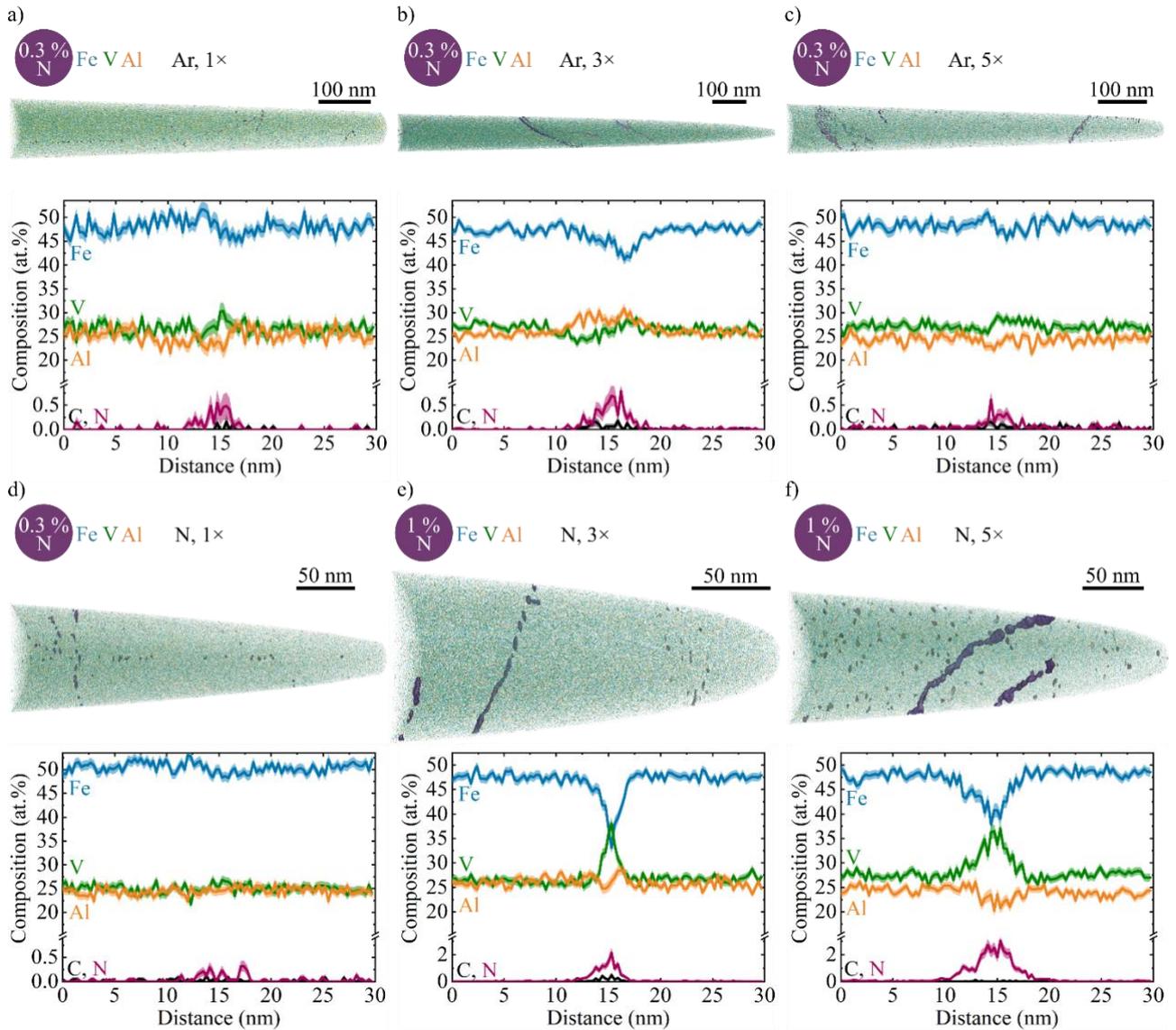

**Figure 7**: APT reconstruction of specimens, selected to analyse dislocations, observed due to segregation. The segregation is visualized by iso-composition surfaces with a threshold given in the figure. The compositional profiles are calculated using a cuboidal region-of-interest with a size of (30 × 5 × 30) nm³. a) LSR$_{Ar,1\times}$, b) LSR$_{Ar,3\times}$, c) LSR$_{Ar,5\times}$, d) LSR$_{N,1\times}$, e) LSR$_{N,3\times}$, f) LSR$_{N,5\times}$.

In addition to individual dislocations within the grains, we observed segregation to dislocation arrays that constitute LAGBs. Fig. 8 shows APT reconstructions of the specimen with LAGBs with the same iso-concentration surfaces as in Fig. 7. Mainly tilt boundaries with parallel dislocations are observed, but a twist boundary with crossed dislocation lines was found in the LSR$_{N,5\times}$ sample. For tilt boundaries, the distance between the dislocations can be used to calculate the misorientation angle (see ESI). Dislocation distances between 8 nm and 26.5 nm are found, translating into misorientations of 1.9° and 0.6°, which fit the EBSD results and confirm that these features are, indeed, dislocations.

1D concentration profiles in Figs. 7 and 8 reveal similar trends regardless of whether dislocations are individual ones or within a LAGB. In addition, the level of segregation found at the dislocations which build up the LAGB is independent of the grain boundary misorientation angle. Hence, the averaged composition of the LAGBs depends strongly on misorientation as the dislocation density increases with misorientation angle. A twist boundary, visible in the LSR$_{N, 5\times}$ sample, shows crossed dislocations. At the crossing points, the N composition is increased up to 6 at.%. We assume that these positions are possible nucleation sites for precipitation.

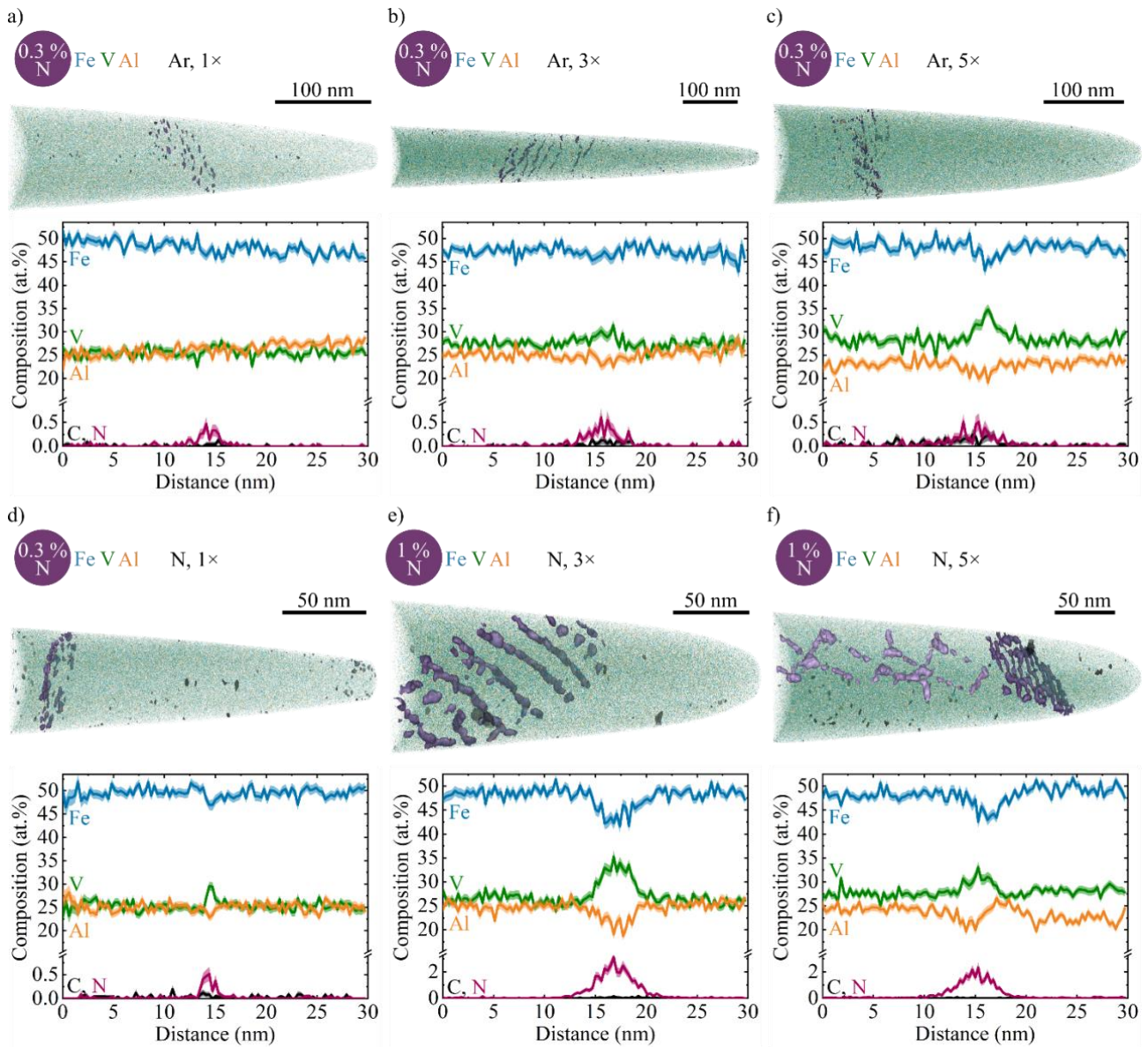

**Figure 8:** APT reconstruction of specimens, selected to analyse grain boundaries, observed due to segregation. The segregation is visualized by iso-composition surfaces with a threshold given in the figure. The compositional profiles are calculated within an individual dislocation using a cuboidal region-of-interest with a size of (30 × 5 × 30) nm³. a) LSR$_{Ar,1\times}$, b) LSR$_{Ar,3\times}$, c) LSR$_{Ar,5\times}$, d) LSR$_{N,1\times}$, e) LSR$_{N,3\times}$, f) LSR$_{N,5\times}$.

Finally, the average nitrogen content of the defects in the different samples is compared. The maximum N composition was determined using the 1D compositional profile being fit to a Gaussian curve. The maximum value was obtained by averaging over 3 to 8 dislocations and 2 to 3 grain boundaries, with 3 to 4 dislocations taken in each boundary for every sample. The overall results are summarized in Fig. 9. As explained before, nitrogen segregation was found in all samples. However, while the composition stays roughly constant for the LSR$_{Ar}$ sample, the N composition found in the LSR$_N$ sample is strongly increased by repeated remelting. For 1× remelting, the composition of defects in the LSR$_N$ sample is similar to that of the LSR$_{Ar}$ samples. For the LSR$_{N,1\times}$ sample, the influence of nitrogen can only be observed at the surface in form of nano-precipitation, as indicated by ECCI measurements. Upon repeated remelting, these precipitates get turbulently mixed into the melt pool. During and after solidification, the inserted nitrogen co-segregates together with vanadium to dislocations and grain boundaries. For the 3× and 5× remelted regions, the N composition is increased by a factor of 4 or 8, respectively. As the interaction time between the atmosphere and the material cannot only be increased by repeated remelting, but also by a reduction of the laser scanning speed, our approach of *in-situ* nitriding offers the possibility to control the nitrogen composition at defects.

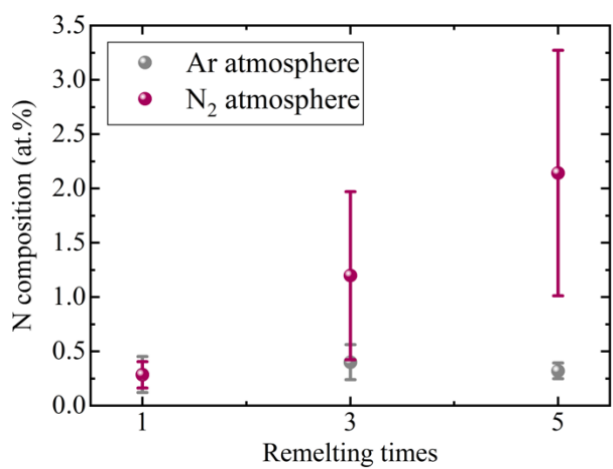

**Figure 9**: Comparison of the N composition observed at dislocations and grain boundaries for LSR$_{Ar}$ (grey) and LSR$_{N}$ (dark red) samples. The error bars show the standard deviation of the calculation.

Conclusions

To summarize, we present an approach to perform laser surface remelting in a reactive N$_2$ atmosphere to form nitrides and induce segregation of nitrogen to crystal defects on the example of thermoelectric full-Heusler compound Fe$_2$VAl. We show that the amount of nitrogen at dislocations and low-angle grain boundaries can be controlled by repeated remelting. An increase of nitrogen content by two orders of magnitude was revealed by atom probe tomography. Nitrogen is co-segregated by vanadium, creating a VN$_x$ Cottrell atmosphere around dislocations. Due to the strains within the melt pool area, a high density of geometrically necessary dislocations in the order of $10^{13}$ m$^{-2}$ is observed by EBSD and visualized by ECCI. Our approach of *in-situ* nitriding during remelting can be applied to other additive manufacturing processes, manipulating physical properties by inducing local segregation and precipitation.

## Author Contributions

L. G.: investigation, data curation, conceptualization, investigation, formal analysis, visualization, writing – original draft; S.-P. T.: data curation, formal analysis, visualization, writing – review & editing; M. R.: investigation, writing – review & editing; R. B. V.: investigation, writing – review & editing; P. K.: Software, writing – review & editing; S. Z.: supervision, writing –review & editing; C. S.: supervision, writing –review & editing; B. G.: resources, supervision, conceptualization, writing – review & editing

## Conflicts of interest

There are no conflicts to declare.

## Acknowledgments

We thank U. Tezins, A. Sturm, M. Nellessen, C. Broß, and K. An-genendt for their technical support at the FIB/APT/SEM facilities at MPIE. LG gratefully acknowledges Studienstiftung des deutschen Volkes for funding. SPT acknowledges Nippon Steel Corporation for financial support. RBV acknowledges the International Max Planck Research School for Interface Controlled Materials for Energy Conversion (IMPRS-SurMat) for funding.